\documentclass[pra,twocolumn,fleqn,balancelastpage]{revtex4-1}%
\pdfoutput=1
\usepackage{amssymb}
\usepackage{amsfonts}
\usepackage{amsmath}
\usepackage{graphicx}%
\providecommand{\U}[1]{\protect\rule{.1in}{.1in}}

\begin{document}
\title{Pattern formation in a complex Swift-Hohenberg equation with phase bistability}
\author{Manuel Mart\'{\i}nez-Quesada and Germ\'{a}n J. de Valc\'{a}rcel}
\affiliation{Departament d'\`{O}ptica, Universitat de Val\`{e}ncia. Dr. Moliner 50 46100
Burjassot, Spain}

\begin{abstract}
We study pattern formation in a complex Swift Hohenberg equation with
phase-sensitive (parametric) gain. Such an equation serves as a universal
order parameter equation describing the onset of spontaneous oscillations in
extended systems submitted to a kind of forcing dubbed rocking when the
instability is towards long wavelengths. Applications include two-level lasers
and photorefractive oscillators. Under rocking, the original continuous phase
symmetry of the system is replaced by a discrete one, so that phase
bistability emerges. This leads to the spontaneous formation of phase-locked
spatial structures like phase domains and dark-ring (phase-) cavity solitons.
Stability of the homogeneous solutions is studied and numerical simulations
are made covering all the dynamical regimes of the model, which turn out to be
very rich. Formal derivations of the rocked complex Swift-Hohenberg equation,
using multiple scale techniques, are given for the two-level laser and the
photorefractive oscillator

\end{abstract}
\maketitle

\section{ Introduction}

The dynamics of nonlinear systems is largely nontrivial and numerical
simulations, involving sophisticated mathematical techniques, usually need to
be made to fully understand their temporal evolution. The situation is even
more complicated in spatially extended systems, like in nonlinear optics, in
which (multi)dimensional variables are present in the dynamics. In these cases
an analytical (or semi-analytical) approach that allows us to gain physical
insight about the system is only possible if we consider universal models
\cite{Cross}, which formally capture the dynamics of nonlinear systems close
to critical points (like the threshold of emission in a laser). These models,
also known as order parameter equations (OPE) provide a simplified yet
powerful description of the system. Their universal character is due to the
fact that very different systems (biological, chemical, physical, etc.) are
described by the same OPE, the only difference being in the meaning of
variables and parameters under this approach.

The symmetries of the system play a central role in the final form of the OPE,
those equations are able to capture qualitatively the essential elements of
the dynamics. These OPE can be real or complex depending on the fundamental
variable (electric field, temperature) that is considered. In nonlinear
optics, complex OPE are commonly used since a complex Ginzburg Landau (CGLE)
for lasers for finite positive detuning was derived \cite{cgr} and have proven
being very helpful for understanding a variety of systems \cite{victorlibro}.
Another equation, a complex Swift-Hohenberg (CSHE), which is valid for small
detunings (positive and negative) was later obtained \cite{lmn}.

One fundamental symmetry for complex OPE describing spatially extended
nonlinear systems is the phase symmetry, as it determines the nature of
patterns which are possible in them. In systems with continuous phase
symmetry, the dynamics is usually turbulent and the presence of vortices and
spiral waves is common. When the systems only allows a finite number of phases
(by means a $n:m$ forcing \cite{coullet2} for instance) the dynamics become
more ordered and controllable. Furthermore, when a $2:1$ (parametric) forcing
is applied to a systems with continuous phase symmetry , the dynamics of the
system allows only two phases in it, which leads to the appearance of novel
structures in the system like phase domains, domain walls and localized
structures (cavity solitons). The proper OPE for the system must reflect this
change and a generic equation can be derived for systems with $n:m$ forcing
\cite{coullet2}. For parametric forcing a family of OPE, like the parametric
complex Ginzburg Landau (PCGLE) \cite{coullet1} are obtained.

Optical systems (like lasers) are usually insensitive to parametric forcing as
the nonlinear response of these systems to high frequencies (twice the natural
frequency) is negligible. In the last years a novel technique involving a new
kind of forcing, known as "rocking" , was introduced \cite{germanrock1} . This
is a $1:1$ forcing (the frequency of the injection is close to the frequency
of the system) so it is appropriate for optical systems. The key factor is
that the injection oscillates in time (temporal rocking) \cite{germanrock1} or
space (spatial rocking) \cite{germanrock2} with a given frequency. This
modulation modifies the dynamics of the system which becomes phase bistable.
It can be shown that a PCGLE describes the universal dynamics of a laser close
to threshold under rocking when the undriven system exhibits an homogeneous
Hopf bifurcation.

Rocking has been successfully applied (theoretically and experimentally) to a
wide range of systems \cite{germanrock3}. This has led to the derivation of
OPEs describing those systems under certain limits and that provide relevant
information of the dynamics. Numerical simulations also prove that the
behaviour predicted by these OPEs extends (qualitatively) far beyond the
conditions imposed in their derivation, which increases the utility of this
universal description. A complex Swift-Hohenberg equation with parametric
gain, which describes a photorefractive oscillator (PRO) under the injection
of rocking was derived in \cite{adolfo1}. Here we show that a very similar
equation describes lasers with small detuning when rocking forcing is present,
as well as we generalize the result of \ \cite{adolfo1} to more general setups.

The structure of the article is as follows. In section II we present the
equation which is object of our study and obtain their homogeneous solutions.
In section III a linear stability analysis is performed to both trivial and
homogeneous solutions to study instabilities against perturbations of
wavevector $k$. In section IV we present some numerical simulations of our
model for different values of parameters and show the patterns that can be
found. We conclude in section V. In the appendices we present the derivation
of the equation for lasers and PRO and a comparison with the real
Swift-Hohenberg equation.

\section{The model}

We consider the spatio-temporal dynamics of extended systems close to a
bifurcation to traveling waves of long wavelength, when the system is forced
in time close to its natural frequency (1:1 resonance) and uniformly in space.
However the type of forcing considered here is not the classic periodic one
but has an amplitude which varies on time, specifically whose sign alternates
periodically in time. The kind of forcing which we refer to can be thus
expressed as:%
\begin{equation}
F(t)=P(t)e^{-i\omega_{R}t}+c.c. \label{forcing}%
\end{equation}
where $P(t+T)=P(t)$ is a $T$-periodic function of time of period $T\ll
\frac{2\pi}{\omega_{R}}$, and $\omega_{R}$ is a high frequency, almost
resonant with the natural frequency of the undriven system. A simple
realization of $P(t)$ is the function:%
\begin{equation}
P(t)=F_{0}\cos(\omega t), \label{periodic}%
\end{equation}
where $\omega\ll\omega_{R}$ and $F_{0}$ is an amplitude.\ According to
(\ref{forcing}) and (\ref{periodic}) the forcing phase (sign) alternates in time.

As shown in the Appendices, where two relevant optical systems are analyzed
under the action of such forcing, the state of the system can be expressed as:%
\[
A\left(  \mathbf{r},t\right)  =A_{R}\left(  t\right)  +\psi\left(
\mathbf{r},t\right)  ,
\]
where $A_{R}(t)$ is a $T$-periodic, spatially uniform contribution merely
following the forcing, plus a spatially 2-dimensional field $\psi\left(
\mathbf{r},t\right)  $ whose dynamics is governed by the following
parametrically driven Swift-Hohenberg equation (written in dimensionless
form):%
\begin{gather}
\partial_{t}\psi\left(  \mathbf{r},t\right)  =\mu(1+i\beta)(1-\left\vert
\psi\right\vert ^{2})\psi+i\alpha\nabla^{2}\psi\nonumber\\
-\left(  \Delta-\nabla^{2}\right)  ^{2}\psi-i\theta\psi+\gamma(1+i\beta
)(\psi^{\ast}-2\psi). \label{swe1}%
\end{gather}
The first line of this equation contains the usual complex Swift-Hohenberg
equation, which models pattern formation arising from a finite wave number
instability to traveling waves close to threshold
\cite{lmn,kestas1,kestas2,longhi1,longhi2,victor0}, whereas in the second line
we find the additional terms which appear when rocking is considered
\cite{germanrock1,germanrock2,germanrock3,adolfo1}. The first extra term,
$-i\theta\psi$, merely moves the reference frame to the central frequency of
the rocking $\omega_{R}$ (see Appendices). The last extra term, $\gamma
(1+i\beta)\psi^{\ast}$, is the actual novelty as it breaks the phase
invariance of the system, which becomes phase bistable as the equation only
has the discrete symmetry $\psi\rightarrow-\psi$. As for the parameters, $\mu$
measures the distance from threshold (it can be removed from the equation by
simple scaling but we keep it not to overwhelm the notation), $\beta$ controls
the nonlinear frequency shift of the system, $\alpha$ controls
diffraction/dispersion. Other parameters are $\Delta$, the detuning of the
cavity from the natural frequency of the unforced system in the optical case,
and $\theta$, which is the detuning of the forcing from the natural frequency
of the system. Finally the "rocking parameter" $\gamma$ is proportional to the
squared amplitude of rocking $F_{0}^{2}$ and also depends on its frequency
$\omega$ in a way whose exact form depends on the system considered (see
Appendices); note that when $\gamma=0$ the effect of rocking is lost and
(\ref{swe1}) becomes a usual complex Swift-Hohenberg equation.

In the following we will take $\beta=0$ as to not increase the degrees of
freedom of the problem without need and also because in the analyzed cases
that parameter happens to be zero. On the other hand $\alpha=1$ in the PRO
case, while $\alpha\geq1$ in the case of laser.

\section{Phase-locked spatially uniform solutions: Rocked states and their
stability}

The spatially uniform nontrivial solutions of (\ref{swe1}), or "rocked
states", can be expressed as $\psi_{\pm}=\left\vert \psi_{\pm}\right\vert
e^{i\phi_{\pm}}$ where:%
\begin{subequations}
\begin{align}
\left\vert \psi_{\pm}\right\vert ^{2}  &  =\mu-2\gamma-\Delta^{2}\pm
\sqrt{\gamma^{2}-\theta^{2}}\label{rock}\\
e^{-2i\phi_{\pm}}  &  =-\frac{i\theta\pm\sqrt{\gamma^{2}-\theta^{2}}}{\gamma}.
\end{align}
\qquad\qquad\ 

The state $\psi_{-}$ is always unstable as follows from a standard linear
stability analysis, so we will not consider it in the following. On the other
hand as $\phi_{+}$ can take two values (differing by $\pi$) this produces two
phase-locked states (with same amplitude) and phase differing by $\pi$
(bistable phase locking). From now on we rename $\psi_{+}\equiv\psi_{0}$
($\phi_{+}=\phi_{0}$). The existence of these states requires $\mu-\Delta
^{2}>0$ and $\gamma>\left\vert \theta\right\vert $; see (\ref{rock}).
\ Moreover, they exist only if $\gamma_{0}<\gamma<\gamma_{+}$, where:%
\end{subequations}
\begin{align}
\gamma_{0}  &  =\left\{
\begin{array}
[c]{c}%
\left\vert \theta\right\vert ,\text{ if }\left\vert \theta\right\vert
<(\mu-\Delta^{2})/2\\
\gamma_{-}\text{ otherwise}%
\end{array}
\right.  \text{ \ \ \ \ \ \ }\nonumber\\
\text{\ }\gamma_{\pm}  &  =\frac{2(\mu-\theta^{2})}{3}\text{\ \ }\pm\frac
{1}{3}\sqrt{(\mu-\theta^{2})^{2}-3\Delta^{2}}.
\end{align}

On the plane $\gamma-\theta$ the existence region is a closed domain ("rocking
balloon") as we can see in Fig. 1. Outside that region we have the trivial
solution and phase unlocked solutions (traveling waves or patterns).%
\begin{figure}
[ptb]
\begin{center}
\includegraphics[
natheight=18.777700in,
natwidth=13.888900in,
height=3.7836in,
width=2.8046in
]%
{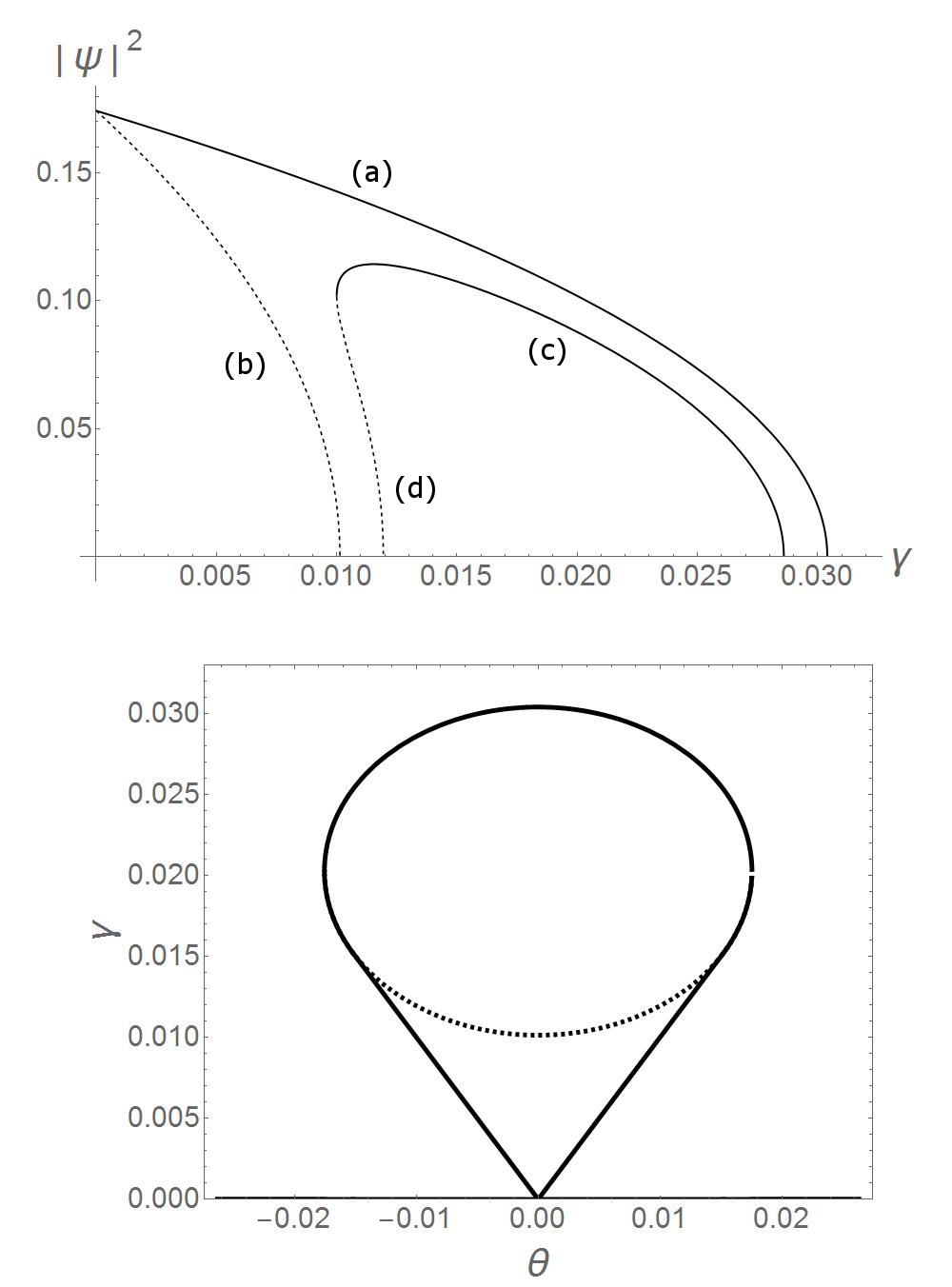}%
\caption{(Top) Amplitude of the rocking states versus $\gamma$ for $\theta=0$
, (a) $\psi_{+}$ ,\ (b) $\psi_{-}$ and $\theta=0.01$ , (c) $\psi_{+}$ ,\ (d)
$\psi_{-}$ (Bottom) Domain of existence of $\psi_{+}$ (bold line)\ and
$\psi_{-}$ (dotted line) in the $\theta-$ $\gamma$ plane; the lower bound is
the same for both functions.Rest of parameters are $\mu=0.05,$ $\Delta=0.14,$
$\alpha=2.$}%
\label{Fig1}%
\end{center}
\end{figure}

We have performed a linear stability analysis of the trivial and uniform
solutions against perturbations with wavevector $k$. The eigenvalue with the
largest real part (for the stability of the trivial solution we just set
$\psi_{0}=0$) reads:%
\begin{gather}
\lambda\left(  k\right)  =\mu-2\gamma-\left\vert \psi_{0}\right\vert
^{2}-\left(  k^{2}+\Delta\right)  ^{2}+\nonumber\\
\sqrt{\left\vert \psi_{0}\right\vert ^{4}-2\gamma\cos\left(  2\phi_{0}\right)
\left\vert \psi_{0}\right\vert ^{2}-\left[  \left(  \alpha k^{2}%
+\theta\right)  ^{2}-\gamma^{2}\right]  } \label{eigenval}%
\end{gather}

As we have two different nonlocal terms in the Swift-Hohenberg (\ref{swe1}),
$\lambda(k)$ will have, in general, two local maxima. One of them $(k_{s})$ is
associated with a real eigenvalue and it will give rise to static patterns
through a patten-forming bifurcation; the another one $(k_{o})$ corresponds to
a complex eigenvalue and it will produce oscillatory solutions (homogeneous or
traveling waves) through a Hopf bifurcation from the trivial solution. We
could not obtain an exact analytical expression for $k_{R}$ but an
approximated one by considering (as it is observed numerically) that it is
close to $-\theta/\alpha$. Writing $k=-\theta/\alpha+\varepsilon$ in
(\ref{eigenval}) and expanding $\lambda$ to second order in $\varepsilon$, the
resulting quadratic expression can\ be maximized and solved for $\varepsilon$,
leading to%
\begin{equation}
k_{s}^{2}=\max\left(  -\frac{\alpha\theta+2\Delta S}{\alpha^{2}+2S},0\right)
, \label{realeigen}%
\end{equation}
where $S=\gamma$ for the trivial solution and $S=\sqrt{\theta^{2}%
+(2\gamma+\Delta^{2}-\mu)^{2}}$ in the case of the rocked states. The
expression for $k_{o}^{2}$ is just
\[
k_{o}^{2}=\max(-\Delta,0),
\]
so for $\Delta>0$ we will have homogenous oscillations while for $\Delta<0$ we
will obtain traveling waves.%
\begin{figure}
[ptb]
\begin{center}
\includegraphics[
natheight=12.260500in,
natwidth=19.635599in,
height=1.99in,
width=3.1689in
]%
{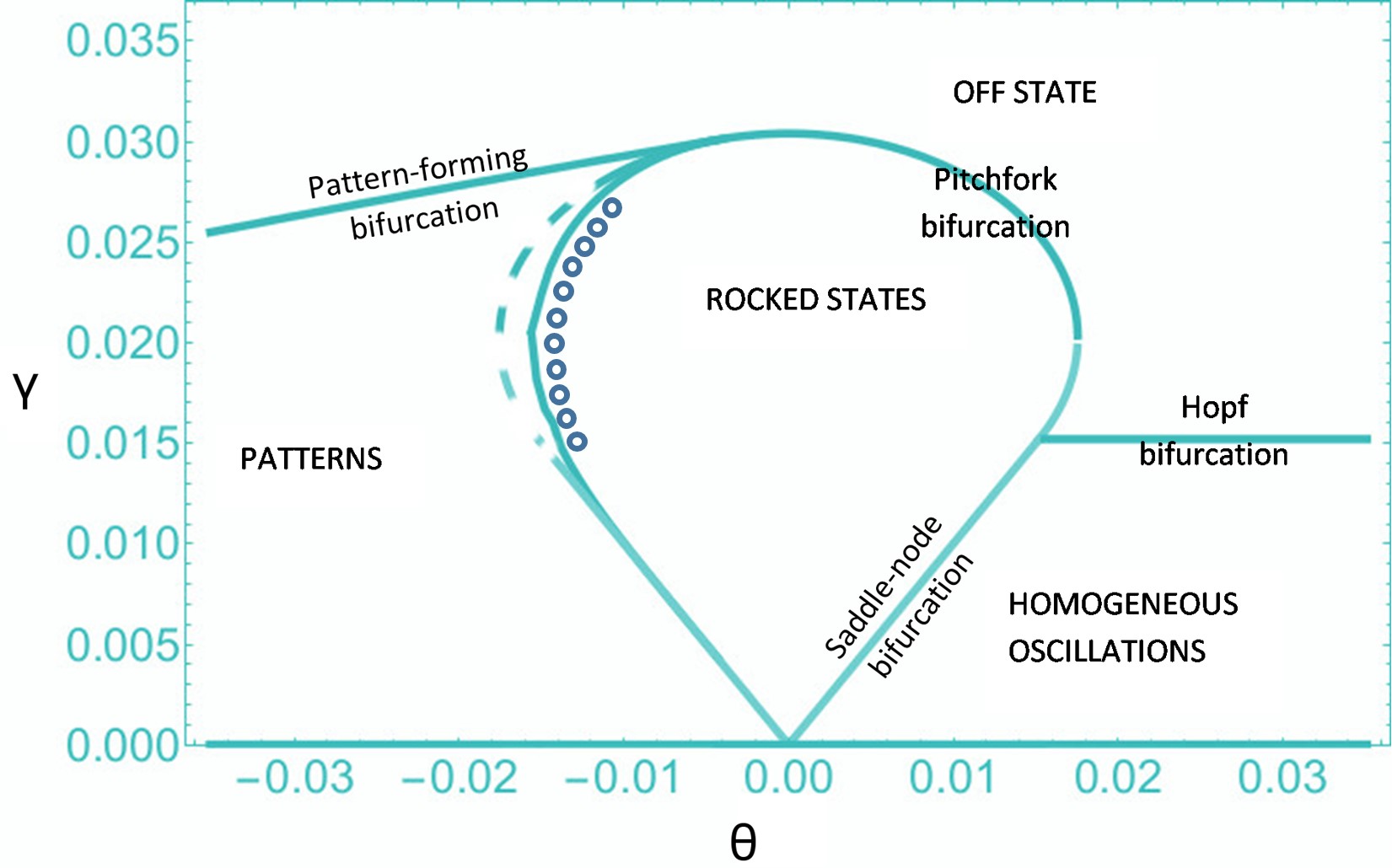}%
\caption{Bifurcation diagram of equation (\ref{swe1}) for $\alpha
=2,\mu=0.05,\Delta=0.14.$ The dashed line indicates the left boundary of
existence of the uniform rocked states, which are become unstable before
reaching it. The circles refer to the existence of dark-cavity solitons.}%
\label{Fig2}%
\end{center}
\end{figure}

\bigskip

\qquad\ 

All these results are summarized in Fig. 2 (positive $\Delta$) and Fig. 3
(negative $\Delta$) for $\alpha=2$ (laser case) and $\mu=0.05$. For positive
$\Delta,$ if $\gamma$ large, the trivial solution, which always exists, is
stable. As we decrease the rocking parameter $\gamma$, we see bifurcations to
other solutions like static patterns (for the trivial state, $k_{s}%
^{2}>0\Longleftrightarrow\theta<-2\gamma\Delta/\alpha$) and homogeneous
oscillations (for $\gamma=(\mu-\Delta^{2})/2,$ $\theta>\gamma,$ $k_{C}^{2}%
=0)$. The transition to uniform rocked states is supercritical in the upper
bound (the amplitude $\psi_{0}$ becomes $0$ along that line). Regarding the
lower bound (two straight lines, where $\psi_{+}=\psi_{-}$) there is a
saddle-node bifurcation which connects with oscillations by means of a complex
eigenvalue for positive $\theta$ . For negative $\theta$ the rocked states
become unstable close to the left edge of the balloon to instabilities (real
eigenvalue) of wavenumber \ $k_{s}^{2}$. In Fig.4 we can see an example of the
temporal dynamics of spatially homogenous states close to the bifurcation: as
we go closer to this, the period of the oscillations becomes larger and is
infinite at the bifurcation.%
\begin{figure}
[ptb]
\begin{center}
\includegraphics[
natheight=11.510700in,
natwidth=18.417101in,
height=1.8696in,
width=2.9738in
]%
{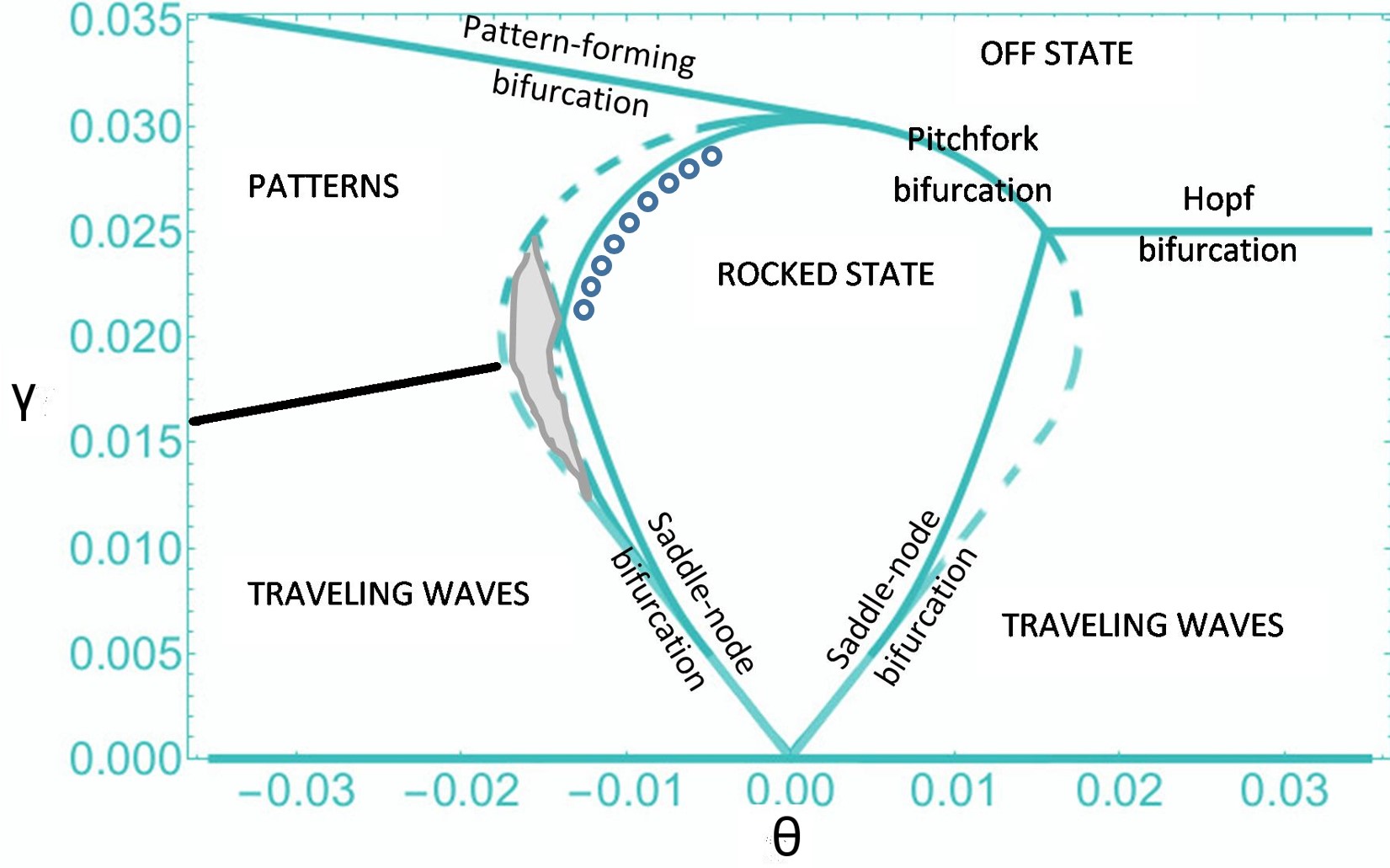}%
\caption{Bifurcation diagram of equation (\ref{swe1}) for $\alpha
=2,\mu=0.05,\Delta=-0.14.$ The dashed line indicates the boundary of existence
of the uniform rocked states, which become unstable before reaching it. In the
shadowed region the two instabilities (with two different spatial frequencies)
for the uniform rocked states are present as explained above. The circles
refer to the existence of dark-cavity solitons.}%
\label{Fig3}%
\end{center}
\end{figure}

\bigskip%
\begin{figure}
[ptb]
\begin{center}
\includegraphics[
natheight=4.555800in,
natwidth=13.888900in,
height=2.6481in,
width=3.3615in
]%
{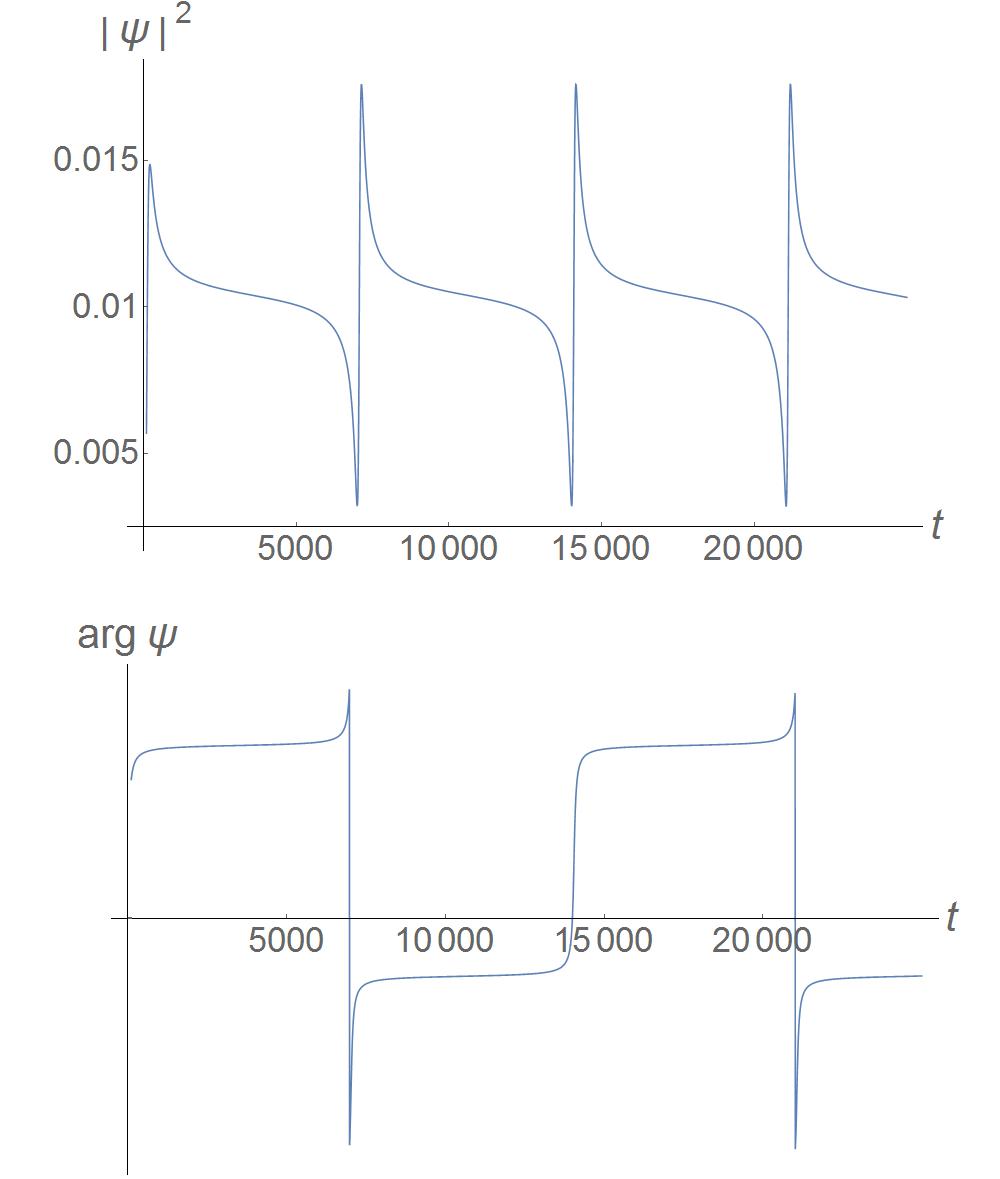}%
\caption{Temporal evolution of intensity (top) and phase (bottom) of a
homogeneous solution close to the saddle-node bifurcation in Fig. 3.
Parameters are $\mu=0.05,$ $\Delta=-0.14,$ $\alpha=2,$ $\gamma=0.01,$
$\theta=0.01001.$}%
\end{center}
\end{figure}

The analysis for negative $\Delta$ is richer, as expected, because in this
regime the Swift-Hohenberg equation (\ref{swe1}) without rocking ($\gamma=0$)
already displays traveling waves of wavenumber $k_{SH}=\sqrt{-\Delta}$). For
large positive $\theta$ we still have a Hopf bifurcation connecting trivial
and traveling wave solutions (for $\gamma=\mu/2$, $\theta>\sqrt{\mu
^{2}-4\Delta^{2}}/2$, $k_{o}^{2}=-\Delta$). For negative, and small positive
$\theta$ ($\theta<-2\gamma\Delta/\alpha$) the trivial solution destabilizes to
a pattern with leading wavenumber $k_{s}^{2}$ at a certain $\gamma
=\gamma(\theta)$. Additionally, for small $\gamma$ we have traveling waves as
in the previous case. Regarding the rocked states, these are destabilized in
two ways: (i) through a complex eigenvalue with most unstable wavevector
$k_{o}^{2}=-\Delta$, which happens symmetrically around $\theta$ (see Fig. 3)
and (ii) through a real eigenvalue with most unstable wavevector equal to
$k_{s(rocked)}^{2}$, which happens, as in the case of positive detuning, close
to the left side of the balloon (here the unstable region is bigger). The
presence of\ these two unstable wavenumbers can be only seen in the
simulations in the transient development of the instabilities (Fig. 5) as only
one of the two spatial frequencies eventually survives. Close to the lower
bound we find (i) and close to the upper bound we get (ii). In a range of
values of $\Delta(\sim-0.14)$ we find that both instabilities arise
simultaneously (the two peaks of $\lambda\left(  k\right)  $ at $k_{o}$
$\ $and $k_{s(rocked)}$ becomes both positive as in Fig. 6: co-dimension-2
point), this happens for a small region in the $\gamma-\theta$ plane (Fig.3).
This behaviour also appears (for negative $\theta$) outside of the region
where rocked states exist along a line which separates static patterns from
traveling waves (Fig. 3).%
\begin{figure}
[ptb]
\begin{center}
\includegraphics[
natheight=5.885100in,
natwidth=5.968900in,
height=3.3235in,
width=3.371in
]%
{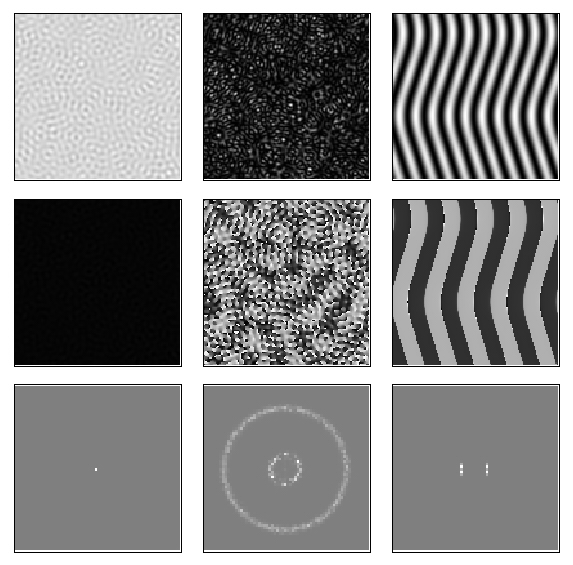}%
\caption{Competition between two instabilities. Snapshots of intensity (upper
row), phase and spatial spectrum (lower row) of the transient dynamics in
(left to right) $t=150$, $t=1725$ and $t=120000$. Paremeters are $\mu=0.05,$
$\Delta=-0.14,$ $\alpha=2,$ $\gamma=0.0297,$ $\theta=-0.0142.$Size of the
windows is 400x400.}%
\label{Fig5}%
\end{center}
\end{figure}
%

\begin{figure}
[ptb]
\begin{center}
\includegraphics[
natheight=4.083600in,
natwidth=7.235900in,
height=1.2514in,
width=2.1983in
]%
{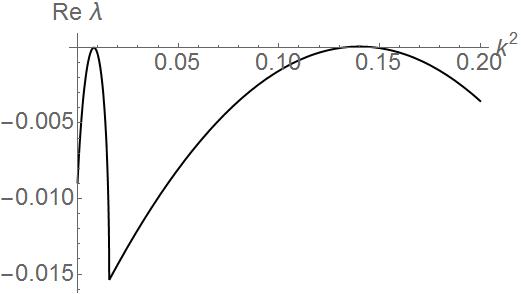}%
\caption{Real part of the eigenvalue $\lambda\left(  k\right)  $ obtained in
the linear stability analysis of the URS solution for parameters: $\mu=0.05,$
$\Delta=-0.14,$ $\alpha=2,$ $\gamma=0.0205,$ $\theta=-0.0139$}%
\label{Fig6}%
\end{center}
\end{figure}

\section{Spatial structures}

We studied the spatial patterns which appear outside the tongue where rocked
(phase-bistable) states exist. In the former case, they basically confirm the
previous analysis as we were able to obtain static spatial patterns which
arise from real eigenvalue instabilities, selecting a particular spatial
frequency as we have seen in the previous analysis. Then the (slow) dynamics
finally leads to a pattern where two spatial modes (of opposite wavenumber)
are dominant, leading to roll-like patterns (as in the right column of Fig.
5). Additionally, we observe phase-unlocked patterns as (slightly modulated)
traveling waves for negative $\Delta$ (Fig. 7), as a result of an instability
governed by a complex eigenvalue.%

\begin{figure}
[ptb]
\begin{center}
\includegraphics[
natheight=2.010700in,
natwidth=5.968900in,
height=1.2332in,
width=3.6097in
]%
{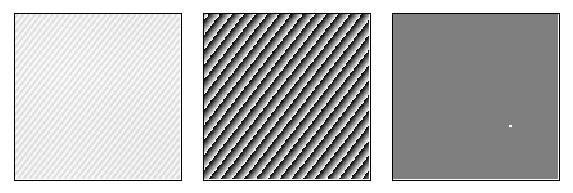}%
\caption{Intensity (left), phase and spatial spectrum (right) of a traveling
wave pattern obtained for $\mu=0.05,$ $\Delta=-0.14,$ $\alpha=2,$
$\gamma=0.012,$ $\theta=-0.025.$Size of the windows is 500x500.}%
\label{Fig7}%
\end{center}
\end{figure}

On the other hand, numerical experiments of (\ref{swe1}) confirm the existence
of a variety of spatial patterns due to the phase bistability and the
instabilities studied in the previous section. In the simulations we fix
$\alpha=2$ (remember that $\alpha=1$ for photorefractive oscillators
and\ $\alpha\geq1$ for lasers) as it can be shown that, after proper
rescaling, the parameter $\alpha$ does not affect the dynamics of the system
and its value can be set an arbitrarily. Different values of $\alpha$ just
change the temporal scale and the spatial scale associated to the real
eigenvalue instability but what is relevant to the dynamics is the ratio
between that scale and the one associated to the complex eigenvalue, which is
determined by $\Delta$.%

\begin{figure}
[ptb]
\begin{center}
\includegraphics[
natheight=3.947900in,
natwidth=5.968900in,
height=2.1975in,
width=3.3105in
]%
{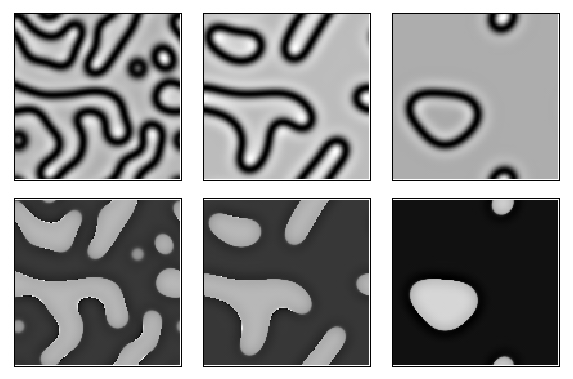}%
\caption{Contracting phase domains. Snapshots of intensity (upper row),
phase(lower row) of transient dynamics of phase domains starting from noise in
(left to right) $t=1000$, $t=3000$ and $t=7500$. Paremeters are $\mu=0.05,$
$\Delta=0.14,$ $\alpha=2,$ $\gamma=0.02,$ $\theta=-0.005.$Rest of parameters
as in Fig.7}%
\label{Fig8}%
\end{center}
\end{figure}

Traveling waves (negative $\Delta$) and homogenous oscillations (positive
$\Delta$) are found in the phase-unlocked regions. Inside the "rocked"
balloon, two uniform rocked states of opposite phase can be connected through
domain walls, generating phase domains (Fig. 8). In two spatial dimensions,
these domains are always a transient state before one phase dominates
\cite{curva}. Close (but still inside the stability region) to the edge where
the uniform rocked states lose their stability with real eigenvalue (see
Section III) the walls become unstable due to curvature effects \cite{curva}
giving rise to the appearance of labyrinth patterns (Fig. 9). Before reaching
the threshold where these labyrinths appear, we find dark-ring cavity solitons
\cite{victorlibro,adolfo1,victor1,victor2,gomila,taranenko}, which can be
written/erased as it is shown in Fig. 10. This happens for both positive and
negative $\Delta.$%

\begin{figure}
[ptb]
\begin{center}
\includegraphics[
natheight=3.947900in,
natwidth=5.968900in,
height=2.1976in,
width=3.3101in
]%
{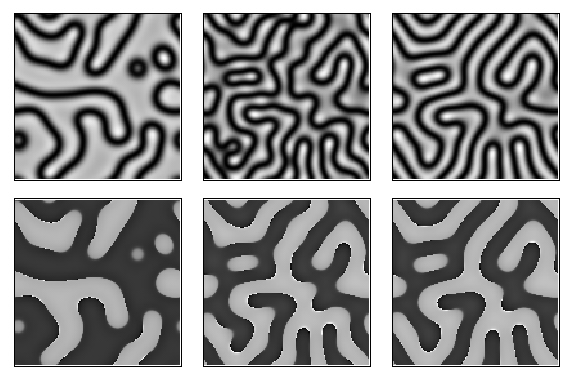}%
\caption{Labyrinth formation. Snapshots of intensity (upper row), phase(lower
row) of transient dynamics of labyrinths starting from left picture in Fig.8
in (left to right) $t=0$, $t=500$ and $t=3000$ for $\theta=-0.012$. Rest of
paremeters as in Fig.8}%
\label{Fig9}%
\end{center}
\end{figure}
%

\begin{figure}
[ptb]
\begin{center}
\includegraphics[
natheight=3.947900in,
natwidth=7.937300in,
height=1.6862in,
width=3.3615in
]%
{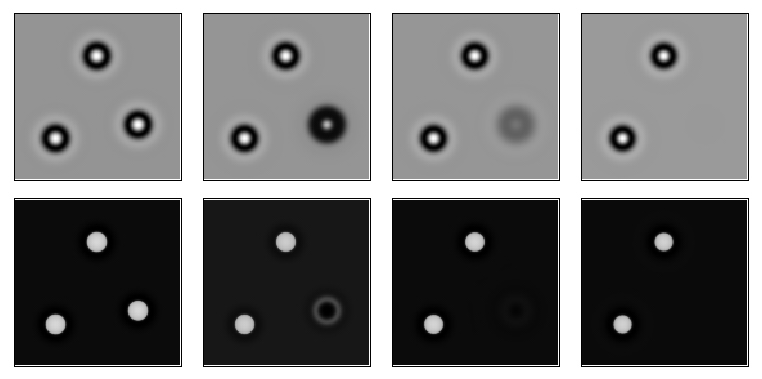}%
\caption{Writing/Erasing phase cavity solitons. Intensity (upper row) and
phase (lower row) plots of transient dynamics of the erasing of a a dark-ring
cavity soliton for $\theta=-0.011$. Times (left to right) are
$t=0,t=150,t=300,t=450$. Rest of parameters as in Fig.8}%
\label{Fig10}%
\end{center}
\end{figure}

\section{Conclusions}

We have proposed and studied, analytically and numerically, a complex
Swift-Hohenberg equation with a parametric term that breaks the phase
invariance, giving rise to phase-bistable patterns. This equation models
diverse nonlinear optical cavities and can be thought of as a universal
equation for rocked, spatially extended systems close to a homogeneous Hopf
bifurcation. In particular we have derived such an equation for two optical
cavities:\ the two-level laser and the photorefractive oscillator. The
structure of the equation, with two nonlocal terms, produces two kinds of
instabilities as revealed by a linear stability analysis of the homogeneous
states. These two differ in the character of the eigenvalue governing the
instability (real or complex) and in the spatial scale generating a complex
variety of spatial patterns, phase locked and phase unlocked. The existence of
extended patterns like Ising domain walls and labyrinths is confirmed with
numerical solutions as well as of dark-ring (phase) solitons.

\bigskip\appendix

\section*{Appendix A:\ Laser}

Our starting point is the set of Maxwell-Boch Equations with a bichromatical
injected signal \cite{germanrock3}:%
\begin{align}
\partial_{t}E  &  =\sigma\left[  1-(1+i\Delta)E+P\right]  +i\nabla^{2}%
E+F\cos(\omega t)e^{i\theta t}\label{MB}\\
\partial_{t}P  &  =-(1-i\Delta)P+(r-N)E\nonumber\\
\partial_{t}N  &  =b\left[  -N+\frac{1}{2}\left(  E^{\ast}P+P^{\ast}E\right)
\right] \nonumber
\end{align}

The complex fields $E$ and $P$ are the scaled envelopes of the electric field
and polarization, $-N$ is proportional to the difference between the
population inversion and its steady value in the absence of lasing.
$\sigma=\kappa/\gamma_{\bot}$ and $b=\gamma_{\Vert}/\gamma_{\bot}$, where
$\kappa$, $\gamma_{\bot}$, and $\gamma_{\Vert}$ are, respectively, the decay
rates of $E$, $P$, and $N$. The transverse Laplacian $\nabla^{2}=\partial
_{x}^{2}+\partial_{y}^{2}$, where the spatial coordinates $\left(  x,y\right)
$ have been normalized so as to make unity the diffraction coefficient, and
$t=\gamma_{\bot}T$ where $T$ is the physical time. $r$ is the pump parameter
and the detuning $\Delta=\left(  \omega_{C}-\omega_{A}\right)  /\left(
\gamma_{\bot}+\kappa\right)  $, being $\omega_{C}$ ($\omega_{A}$) the cavity
(atomic) frequency.

Eqs. (\ref{MB}) have been written in the frequency frame $\omega_{0}=\left(
\gamma_{\bot}\omega_{C}+\kappa\omega_{A}\right)  /\left(  \gamma_{\bot}%
+\kappa\right)  $ of the on-axis, or plane-wave $\left(  \nabla^{2}E=0\right)
$, lasing solution in the absence of injected signal. In particular, this
means that the actual injected field at the entrance face of the amplifying
medium, $\mathcal{E}_{in}$, is of the form:%
\begin{align}
\mathcal{E}_{in}  &  =E_{in}\exp\left(  -i\omega_{0}T\right)  +c.c.
\label{inj}\\
&  =E_{in}\exp\left(  -i\omega_{0}\gamma_{\bot}^{-1}t\right)  +c.c.
\end{align}

\bigskip We will consider two types of scales: Fast and slow.

\subsection*{\bigskip Fast Scales}

Regarding the scales, we will consider that the detuning $\Delta$ and
laplacian $\nabla^{2}$ are $\mathcal{O}\left(  \varepsilon\right)  $ $.$As we
consider class C lasers, $\sigma$ and $b$ are $\mathcal{O}\left(
\varepsilon^{0}\right)  .$ Additionally, $F=\mathcal{O}\left(  \varepsilon
^{2}\right)  $ and $\theta=\mathcal{O}\left(  \varepsilon\right)  .$ Time
scales are $T_{1}=\varepsilon t$ and $T_{2}=\varepsilon^{2}t$ , the pump is
$r=1+\varepsilon^{2}r_{2}$ (we are close to the threshold)$.$ The variables
will be written in the form:%
\begin{equation}
(E,P,N)=(E_{0},P_{0},N_{0})+\varepsilon(E_{1},P_{1},N_{1})+\varepsilon
^{2}(E_{2},P_{2},N_{2})+...
\end{equation}

$\bigskip\mathcal{O}\left(  \varepsilon^{0}\right)  $

At this order \ $E_{0}=P_{0}=N_{0}=0.$

$\mathcal{O}\left(  \varepsilon\right)  $

This is the first nontrivial order and reads:%
\begin{align}
N_{1}  &  =0,\label{N1C1}\\
\mathcal{L}_{0}\left\vert v_{1}\right\rangle  &  =0, \label{L0v1C1}%
\end{align}
where%
\[
\mathcal{L}_{0}=\left(
\begin{array}
[c]{cc}%
-\sigma & \sigma\\
r_{0} & -1
\end{array}
\right)  ,
\]

and we have introduced the notation%
\begin{equation}
\left\vert v_{i}\right\rangle =\left(
\begin{array}
[c]{c}%
E_{i}\\
P_{i}%
\end{array}
\right)  ,\quad i=1,2,3,\ldots\label{vi}%
\end{equation}
Eq. (\ref{L0v1C1}) can be easily solved with the help of the left eigenvectors
of matrix $\mathcal{L}_{0}$:%
\begin{align}
\left\langle \zeta_{1}\right\vert \mathcal{L}_{0}  &  =0\left\langle \zeta
_{1}\right\vert ,\quad\left\langle \zeta_{2}\right\vert \mathcal{L}_{0}%
=\mu\left\langle \zeta_{2}\right\vert \label{eigen}\\
\left\langle \zeta_{1}\right\vert  &  =\left(  1,\sigma\right)  ,\label{z1}\\
\left\langle \zeta_{2}\right\vert  &  =\left(  1,-1\right)  ,\quad\mu=-\left(
1+\sigma\right)  . \label{z2}%
\end{align}
Projecting Eq. (\ref{L0v1C1}) onto $\left\langle \zeta_{1}\right\vert $ we
obtain $0=0$, and projecting onto $\left\langle \zeta_{2}\right\vert $ we
obtain%
\begin{equation}
P_{1}=E_{1}, \label{P1C1}%
\end{equation}
hence%
\begin{equation}
\left\vert v_{1}\right\rangle =E_{1}\left(
\begin{array}
[c]{c}%
1\\
1
\end{array}
\right)  . \label{v1C1}%
\end{equation}

$\mathcal{O}\left(  \varepsilon^{2}\right)  $

Making use of Eqs. () and (), we obtain%
\begin{align}
N_{2}  &  =\left\vert E_{1}\right\vert ^{2},\label{N2C1}\\
\frac{\partial}{\partial T_{1}}\left\vert v_{1}\right\rangle  &
=\mathcal{L}_{0}\left\vert v_{2}\right\rangle +\left\vert g_{2}\right\rangle .
\label{dT1v1C1}%
\end{align}

\bigskip

where\ \ \ \ \ \ \ \ \
\begin{equation}
\left\vert g_{2}\right\rangle =\left(
\begin{array}
[c]{c}%
i\nabla^{2}E_{1}-i\sigma\Delta E_{1}+F_{1}\cos\left(  \omega T_{1}\right)
e^{i\theta T_{2}}\\
i\Delta E_{1}%
\end{array}
\right)
\end{equation}

Projecting Eq. (\ref{dT1v1C1}) onto $\left\langle \zeta_{1}\right\vert $ and
making use of Eq. (\ref{P1C1}) we obtain%
\begin{equation}
\left(  \sigma+1\right)  \frac{\partial E_{1}}{\partial T_{1}}=i\nabla
^{2}E_{1}+F_{1}\cos\left(  \omega T_{1}\right)  e^{i\theta T_{2}}.
\label{dT1E1C1}%
\end{equation}
We solve%
\begin{align}
E_{1}(\overrightarrow{x},T_{1},T_{2})  &  =A_{1}(\overrightarrow{x}%
,T_{1},T_{2})+F_{1}(T_{1},T_{2})\\
F_{1}(T_{1},T_{2})  &  \equiv\frac{F}{\omega}\sin\left(  \omega T_{1}\right)
e^{i\theta T_{2}}\\
\left(  \sigma+1\right)  \frac{\partial A_{1}}{\partial T_{1}}  &
=i\nabla^{2}A_{1} \label{ea1}%
\end{align}

Projecting onto $\left\langle \zeta_{2}\right\vert $ and making use of Eq.
(\ref{dT1E1C1}) we obtain%
\begin{equation}
P_{2}=-\left(  \frac{\partial}{\partial T_{1}}-i\Delta\right)  E_{1}+\left(
1+\sigma\right)  E_{2},
\end{equation}

$\mathcal{O}\left(  \varepsilon^{3}\right)  $

\bigskip%
\begin{equation}
\frac{\partial}{\partial T_{1}}\left\vert v_{2}\right\rangle +\frac{\partial
}{\partial T_{2}}\left\vert v_{1}\right\rangle =\mathcal{L}_{0}\left\vert
v_{3}\right\rangle +\left\vert g_{3}\right\rangle
\end{equation}

where%
\[
\left\vert g_{3}\right\rangle =\left(
\begin{array}
[c]{c}%
0\\
\left(  r_{2}-\left\vert E_{1}\right\vert ^{2}\right)  E_{1}+i\Delta P_{2}%
\end{array}
\right)
\]

Projecting%
\begin{gather}
\sigma\left(  \frac{\partial}{\partial T_{1}}-i\Delta\right)  \left[  \left(
\frac{\partial}{\partial T_{1}}-i\Delta\right)  E_{1}+\left(  1+\sigma\right)
E_{2}\right]  =\nonumber\\
-\left(  1+\sigma\right)  \frac{\partial E_{1}}{\partial T_{2}}+\sigma\left(
r_{2}-\left\vert E_{1}\right\vert ^{2}\right)  E_{1}%
\end{gather}

We can rewrite this as following%
\[
\sigma\left(  \frac{\partial}{\partial T_{1}}-i\Delta\right)  \left(
1+\sigma\right)  E_{2}=g(T_{1},T_{2})
\]

where%
\begin{gather}
g(T_{1},T_{2})=-\sigma\left(  \frac{\partial}{\partial T_{1}}-i\Delta\right)
^{2}A_{1}-\left(  1+\sigma\right)  \frac{\partial A_{1}}{\partial T_{2}%
}\nonumber\\
+\sigma\left(  r_{2}-\left\vert A_{1}\right\vert ^{2}\right)  A_{1}%
+-\sigma\left(  \frac{\partial}{\partial T_{1}}-i\Delta\right)  ^{2}%
F_{1}\nonumber\\
-\left(  1+\sigma\right)  \frac{\partial F_{1}}{\partial T_{2}}+\sigma\left(
r_{2}-\left\vert F_{1}\right\vert ^{2}\right)  F_{1}-2\sigma\left\vert
A_{1}\right\vert ^{2}F_{1}\nonumber\\
-2\sigma\left\vert F_{1}\right\vert ^{2}A_{1}-\sigma F_{1}^{2}A_{1}^{\ast
}+\sigma F_{1}^{\ast}A_{1}^{2}%
\end{gather}

\bigskip The solution can be written formally as:%
\[
E_{2}=\frac{1}{\sigma(1+\sigma)}\left(  e^{i\Delta T_{1}}+%
{\displaystyle\int\limits_{0}^{T_{1}}}
g\left(  T_{1},T_{2}\right)  dT_{1}+A_{2}(T_{2})\right)
\]
To ensure convergence it must be true that:%
\begin{equation}
\lim_{T_{1}\rightarrow\infty}\frac{1}{T_{1}}%
{\displaystyle\int\limits_{0}^{T_{1}}}
g\left(  T_{1},T_{2}\right)  dT_{1}=0
\end{equation}

Taking into account that if $A_{1}$ is homogeneous then it does not depend on
$T_{1}$ the following condition must be fulfilled (we make use of
(\ref{ea1})):%
\begin{gather}
\left(  1+\sigma\right)  \frac{\partial A_{1}}{\partial T_{2}}=-\sigma\left(
\frac{i\nabla^{2}}{1+\sigma}-i\Delta\right)  ^{2}A_{1}\label{ea2}\\
+\sigma\left(  r_{2}-\left\vert A_{1}\right\vert ^{2}\right)  A_{1}%
-\sigma\left(  \frac{F}{\omega}\right)  ^{2}A_{1}-\frac{\sigma}{2}\left(
\frac{F}{\omega}\right)  ^{2}e^{2i\theta T_{2}}A_{1}^{\ast}\nonumber
\end{gather}

in order to avoid divergences (in the case $\nabla^{2}A_{1}=0$)

We finally compute the time derivative $\partial_{t}A_{1}=\left(
\varepsilon\frac{\partial}{\partial T_{1}}+\varepsilon^{2}\frac{\partial
}{\partial T_{2}}\right)  \left(  \varepsilon A_{1}\right)  $ up to third
order in $\varepsilon$ which, making use of (\ref{ea2}) and (\ref{ea1}),
setting $\psi=e^{-i\theta T_{2}}A_{1}$ and rescaling to the original
variables, we obtain:%
\begin{gather}
(1+\sigma)\frac{\partial\psi}{\partial t}=\sigma\left(  \frac{i\nabla^{2}%
}{1+\sigma}-i\Delta\right)  ^{2}\psi-(1+\sigma)i\theta\psi+\nonumber\\
i\nabla^{2}\psi+\sigma\left(  r-1-\left\vert \psi\right\vert ^{2}\right)
\psi-2\gamma\psi+\gamma\psi^{\ast}\label{fast}\\
\text{where }\gamma=\frac{1}{2}\frac{F^{2}\sigma}{\left(  1+\sigma\right)
^{2}\omega^{2}}\nonumber
\end{gather}

\subsection*{Slow Scales}

The scales in this case will be:%

\begin{gather}
r=r_{0}+\varepsilon^{2}r_{2}\nonumber\\
r_{0}=1+\Delta^{2}\varepsilon^{2}\text{ \ \ \ \ \ }\Delta\sim O(\varepsilon
)\nonumber\\
\omega\sim O(\varepsilon^{2})\text{ \ \ \ }F\sim O(\varepsilon^{3})\text{
}\nonumber\\
(E,N,P)\sim(E_{1},P_{1,}N_{1})\varepsilon+(E_{2},P_{2,}N_{2})\varepsilon
^{2}+...
\end{gather}

$O(\varepsilon)$%

\begin{align}
N_{1}  &  =0\\
\emph{L}_{0}\left\vert v_{1}\right\rangle  &  =0\nonumber
\end{align}%
\begin{equation}
\emph{L}_{0}=\left(
\begin{array}
[c]{cc}%
-\sigma & \sigma\\
1 & -1
\end{array}
\right)
\end{equation}

The left-eigenvectors are $\left\langle \xi_{1}\right\vert =(1,\sigma)$ and
$\left\langle \xi_{1}\right\vert =(1,-1)$ with eigenvalues $0$ and
$-(1+\sigma)$ respectively. We can use this to write:%
\begin{align}
\left\langle \xi_{1}\right\vert \emph{L}_{0}\left\vert v_{1}\right\rangle  &
=0\left\langle \xi_{1}\right\vert \left.  v_{1}\right\rangle \nonumber\\
\left\langle \xi_{2}\right\vert \emph{L}_{0}\left\vert v_{1}\right\rangle  &
=-(1+\sigma)\left\langle \xi_{2}\right\vert \left.  v_{1}\right\rangle
\nonumber\\
v_{1}  &  =\left(
\begin{array}
[c]{c}%
E_{1}\\
P_{1}%
\end{array}
\right)
\end{align}

We obtain $E_{1}=P_{1}$

$O(\varepsilon^{2})$%
\begin{align}
N_{2}  &  =\left\vert E_{1}\right\vert ^{2}\\
\frac{\partial}{\partial T_{1}}\left\vert v_{1}\right\rangle  &  =\emph{L}%
_{0}\left\vert v_{2}\right\rangle +\left\vert g_{2}\right\rangle \nonumber
\end{align}

\begin{equation}
\left\vert g_{2}\right\rangle =\left(
\begin{array}
[c]{c}%
i\nabla^{2}E_{1}-i\sigma\Delta E_{1}\\
i\Delta E_{1}%
\end{array}
\right)
\end{equation}

Using the same procedure as above, we obtain (if we set $E_{2}=0$):%

\begin{gather}
\left(  1+\sigma\right)  \frac{\partial E_{1}}{\partial T_{1}}=i\nabla
^{2}E_{1}\\
P_{2}=-\frac{i}{\left(  1+\sigma\right)  }\left(  \nabla^{2}E_{1}-\left(
1+\sigma\right)  \Delta E_{1}\right)  \label{p2}%
\end{gather}

$O(\varepsilon^{3})$%
\begin{gather}
\frac{\partial}{\partial T_{1}}\left\vert v_{2}\right\rangle +\frac{\partial
}{\partial T_{2}}\left\vert v_{1}\right\rangle =\emph{L}_{0}\left\vert
v_{3}\right\rangle +\left\vert g_{3}\right\rangle \\
\left\vert g_{3}\right\rangle =\left(
\begin{array}
[c]{c}%
F\cos\left(  \omega T_{1}\right)  e^{i\theta T_{2}}\\
(r_{2}-\left\vert E_{1}\right\vert ^{2}E_{1}+i\Delta P_{2}%
\end{array}
\right)
\end{gather}

Now we obtain the following equation:%
\begin{gather}
\sigma\left(  \frac{\partial}{\partial T_{1}}-i\Delta\right)  P_{2}%
=-(1+\sigma)\frac{\partial E_{1}}{\partial T_{2}}+\nonumber\\
F\cos\left(  \omega T_{1}\right)  e^{i\theta T_{2}}+\sigma(r_{2}-\left\vert
E_{1}\right\vert ^{2})E_{1}%
\end{gather}
\bigskip Using (\ref{p2}) and undoing the scaling:%
\begin{equation}
\frac{\partial E}{\partial t}=\frac{\partial E_{1}}{\partial T_{1}}%
\varepsilon+\frac{\partial E_{1}}{\partial T_{2}}\varepsilon^{2}%
\end{equation}

We can write:%
\begin{gather}
(1+\sigma)\frac{\partial E}{\partial t}=\sigma\left(  \frac{i\nabla^{2}%
}{1+\sigma}-i\Delta\right)  ^{2}E+i\nabla^{2}E+\nonumber\\
\sigma\left(  r-1-\left\vert E\right\vert ^{2}\right)  E+F\cos\left(  \omega
t\right)  e^{i\theta t}%
\end{gather}

Setting $E\equiv Ae^{-i\theta t}$ the previous equation becomes%
\begin{gather}
(1+\sigma)\frac{\partial A}{\partial t}=\sigma\left(  \frac{i\nabla^{2}%
}{1+\sigma}-i\Delta\right)  ^{2}A-(1+\sigma)i\theta A+\nonumber\\
i\nabla^{2}A+\sigma\left(  r-1-\left\vert A\right\vert ^{2}\right)
A+F\cos\left(  \omega t\right)
\end{gather}

If the frequency $\omega$ is high compared with the dynamics of the system we
can set: $T=\omega t$ $\rightarrow\varepsilon^{-1}t$ with $T\gg t$ and as it
is done in \cite{germanrock1} separate the slow dynamics from the fast dynamics:

Specifically $t=\varepsilon^{-1}T+\tau$%
\begin{equation}
A(\tau,T)=A_{0}(\tau,T)+\varepsilon A_{1}(\tau,T)
\end{equation}
\qquad

No we solve at different orders

$O(\varepsilon^{-1})$%
\begin{equation}
(1+\sigma)\frac{\partial A_{0}}{\partial T}=F\cos\left(  T\right)  \rightarrow
A_{0}(T)=\frac{F}{(1+\sigma)\omega}\sin(T)+i\psi(\tau)
\end{equation}

$O(\varepsilon^{0})$

The equation reads at this order:%
\begin{gather}
(1+\sigma)\frac{\partial A_{1}}{\partial T}+(1+\sigma)\frac{\partial\psi
}{\partial\tau}=\nonumber\\
\sigma\left(  \frac{i\nabla^{2}}{1+\sigma}-i\Delta\right)  ^{2}A_{0}%
+(1+\sigma)i\theta A_{0}+i\nabla^{2}A_{0}\nonumber\\
+\sigma\left(  r-1-\left\vert A_{0}\right\vert ^{2}\right)  A_{0}+F\cos\left(
T\right)
\end{gather}

This can be written as $(1+\sigma)\frac{\partial A_{1}(\tau,T)}{\partial
T}=g(T,\tau)$ which can be solved: $A_{1}(\tau,T)=(1+\sigma)^{-1}%
{\displaystyle\int\nolimits_{0}^{T}}
dT^{\prime}g(T^{\prime},\tau)+B(\tau)$. The boundness of this requires:%
\begin{equation}
\lim\frac{1}{T}%
{\displaystyle\int\nolimits_{0}^{T}}
dT^{\prime}g(T^{\prime},\tau)=0
\end{equation}

Therein, we obtain the following solvability condition:%
\begin{gather}
(1+\sigma)\frac{\partial\psi}{\partial t}=\sigma\left(  \frac{i\nabla^{2}%
}{1+\sigma}-i\Delta\right)  ^{2}\psi-(1+\sigma)i\theta\psi+\nonumber\\
i\nabla^{2}\psi+\sigma\left(  r-1-\left\vert \psi\right\vert ^{2}\right)
\psi-2\gamma\psi+\gamma\psi^{\ast}\\
\text{where }\gamma=\frac{1}{2}\frac{F^{2}\sigma}{\left(  1+\sigma\right)
^{2}\omega^{2}}\nonumber
\end{gather}

\bigskip This condition is exactly the same equation as \ref{fast}$,$so we
recover the result that we obtained considering fast scales by just assuming
that $F$ and $\omega$ are large. So, independently of the set of scales we
consider, the final equation is consistent.

Setting $\frac{\nabla^{2}}{1+\sigma}\equiv\nabla^{\prime2}$ , $\tau\equiv
\frac{\sigma}{\sigma+1}t$ , $\theta^{\prime}\equiv\frac{\sigma+1}{\sigma
}\theta$ , $\gamma^{\prime}\equiv\frac{\gamma}{\sigma}$ , $\alpha\equiv
\frac{\sigma+1}{\sigma}$ and defining $\mu=r-1$ we can write (removing the
commas for simplicity):%
\begin{equation}
\partial_{t}\psi=(\mu-2\gamma-i\theta)\psi-\left\vert \psi\right\vert ^{2}%
\psi-\left(  \Delta-\nabla^{2}\right)  ^{2}\psi+i\alpha\nabla^{2}\psi
+\gamma\psi^{\ast}%
\end{equation}

\section*{Appendix B: PRO}

Our starting point is the set of equations as in \cite{adolfo1} in which we
consider a bichromatical injection and we make the change ($E^{\prime
},N^{\prime})\longrightarrow(E,N)e^{i\Delta t}$ for convenience. After
removing the commas, we get:%
\begin{align}
\sigma^{-1}\partial_{t}E  &  =-(1+i\Delta)E+i\nabla^{2}E+N+F\cos(\omega
t)e^{i\theta t}\label{pro}\\
\partial_{t}N  &  =-(1-i\Delta)N+g\frac{E}{1+\left\vert E\right\vert ^{2}%
}\nonumber
\end{align}

where $E(\mathbf{r},t)$ \ is the slowly varying envelope of the intracavity
field, $N(\mathbf{r},t)$ is the complex amplitude of the photorefractive
nonlinear grating, \textbf{r}=(x,y) are the transverse coordinates,
$\nabla^{2}=\partial_{x}^{2}+\partial_{y}^{2},\sigma=\kappa_{P}\tau$ is the
product of the cavity linewidth $\kappa_{P}$ with the photorefractive response
time $\tau$ ($\sigma\gtrsim10^{8}$ under typical conditions, and $\tau\sim1$
s), $t$ is time measured in units of $\tau$, the detuning ($\omega_{C}%
-\omega_{P})/\kappa_{P}$ ($\omega_{P}$ and $\omega_{C}$ are the frequencies of
the pump and its nearest longitudinal mode, respectively), $g$ is the (real)
gain parameter that depends of the crystal parameters and the geometry of the interaction.

The scales will be $\sigma\sim O(\varepsilon^{-4})$ (results are independent
of the specific scale provided that $\sigma$ very large) and $\Delta
,\theta\sim O(\varepsilon),$ which implies $\nabla^{2}\sim O(\varepsilon)$ .
We also assume that $g$ is close to threshold, $g=1+g_{2}\varepsilon^{2}$ and
that $F\sim O(\varepsilon),\omega\sim O(1).$ We will have "two times" , fast
and slow: $t=T_{1}+\varepsilon^{-1}T_{2}$ so that $\partial_{t}=\partial
_{T_{1}}+\varepsilon\partial_{T_{2}}.$We also set:%
\begin{gather}
(E,N)=(E_{0},N_{0})+\varepsilon(E_{1},N_{1})+\varepsilon^{2}(E_{2}%
,N_{2})+...\\
F=\varepsilon F_{1}\cos(\omega T_{1})e^{i\theta T_{2}}%
\end{gather}

Now we will solve the previous equations (\ref{pro}) at different orders. At
order $O(1)$ we trivially have $E_{0}=N_{0}=0.$

$O(\varepsilon)$%
\begin{gather}
0=-E_{1}+N_{1}+F_{1}\cos(\omega T_{1})e^{i\theta T_{2}}\nonumber\\
\partial_{T_{1}}N_{1}=-N_{1}+E_{1}%
\end{gather}

The solution is%
\begin{align*}
N_{1}  &  =E_{1}-F_{1}\cos(\omega T_{1})e^{i\theta T_{2}}\\
E_{1}  &  =F_{1}\cos(\omega T_{1})e^{i\theta T_{2}}+F_{1}/\omega\sin(\omega
T_{1})e^{i\theta T_{2}}\\
+\varphi_{1}(\mathbf{r},T_{2})  &  \equiv E_{1}^{\omega}+\varphi_{1}%
\end{align*}

$O(\varepsilon^{2})$%
\begin{gather}
0=-E_{2}+(-i\Delta+i\nabla^{2})E_{1}+N_{2}\nonumber\\
\partial_{T_{1}}N_{2}+\partial_{T_{2}}N_{1}=-N_{2}+i\Delta N_{1}+E_{2}%
\end{gather}

We solve%
\begin{gather}
N_{2}=E_{2}-(-i\Delta+i\nabla^{2})E_{1}\nonumber\\
E_{2}=\int_{0}^{T_{1}}G_{2}(T_{1},T_{2})dT_{1}+\varphi_{2}(\overrightarrow
{x},T_{2})\equiv E_{2}^{\omega}+\varphi_{2}\nonumber\\
G_{2}(T_{1},T_{2})=(-i\Delta)(-F_{1}\omega\cos(\omega T_{1})e^{i\theta T_{2}%
}+\label{ggrande}\\
F_{1}\cos(\omega T_{1})e^{i\theta T_{2}}+F_{1}\cos(\omega T_{1})e^{i\theta
T_{2}})\nonumber\\
+(i\theta)F_{1}/\omega\sin(\omega T_{1})e^{i\theta T_{2}}-\partial_{T_{2}%
}\varphi_{1}+i\nabla^{2}\varphi_{1}\nonumber
\end{gather}

Solvability of $E_{2}$ requires:%

\begin{equation}
\lim_{T_{1}\rightarrow\infty}\frac{1}{T_{1}}%
{\displaystyle\int\limits_{0}^{T_{1}}}
G_{2}(T_{1},T_{2})dT_{1}=0
\end{equation}

The oscillatory terms in (\ref{ggrande}) vanish so the previous condition
remains:\qquad\qquad%
\begin{equation}
\partial_{T_{2}}\varphi_{1}=i\nabla^{2}\varphi_{1}%
\end{equation}

$O(\varepsilon^{3})$%
\begin{gather}
0=-E_{3}+(-i\Delta+i\nabla^{2})E_{2}+N_{3}\nonumber\\
\partial_{T_{1}}N_{3}+\partial_{T_{2}}N_{2}=-N_{3}+i\Delta N_{2}+\nonumber\\
E_{3}+(g-1)E_{1}+\left\vert E_{1}\right\vert ^{2}E_{1}%
\end{gather}

We already know that%
\begin{align}
E_{1}  &  =E_{1}^{\omega}(T_{1},T_{2})+\varphi_{1}(\mathbf{r},T_{2}%
)\nonumber\\
E_{2}  &  =E_{2}^{\omega}(T_{1},T_{2})+\varphi_{2}(\mathbf{r},T_{2})
\end{align}

We solve:%
\begin{gather}
N_{3}=E_{3}-(-i\Delta+i\nabla^{2})E_{2}\\
E_{3}=\int_{0}^{T_{1}}G_{3}(T_{1},T_{2})dT_{1}+\varphi_{3}(\mathbf{r},T_{2})\\
G_{3}(T_{1},T_{2})=\label{ggrande2}\\
\left[  (-i\Delta)(\partial_{T_{1}}E_{2}^{\omega}+\partial_{T_{2}}%
E_{1}^{\omega})+(g_{2}-1-\Delta^{2})E_{1}^{\omega}-\partial_{T_{2}}%
E_{2}^{\omega}\right]  +\nonumber\\
\left[  -\partial_{T_{2}}\varphi_{2}+(g_{2}-1)\varphi_{1}-(-i\Delta
+i\nabla^{2})^{2}\varphi_{1}+i\nabla^{2}\varphi_{2}-\left\vert \varphi
_{1}\right\vert ^{2}\varphi_{1}\right]  -\nonumber\\
\left[  \left\vert E_{1}^{\omega}\right\vert ^{2}E_{1}^{\omega}+2E_{1}%
^{\omega}\left\vert \varphi_{1}\right\vert ^{2}+(E_{1}^{\omega})^{\ast}%
\varphi_{1}^{2}+2\left\vert E_{1}^{\omega}\right\vert ^{2}\varphi
_{1}+\left\vert E_{1}^{\omega}\right\vert ^{2}\varphi_{1}^{\ast}\right]
\nonumber
\end{gather}

Solvability of $E_{3}$ requires:%
\begin{equation}
\lim_{T_{1}\rightarrow\infty}\frac{1}{T_{1}}%
{\displaystyle\int\limits_{0}^{T_{1}}}
G_{3}(T_{1},T_{2})dT_{1}=0
\end{equation}

The oscillatory terms in (\ref{ggrande2}) vanish so the previous condition
remains:\qquad%
\begin{align}
\partial_{T_{2}}\varphi_{2}  &  =-(-i\Delta+i\nabla^{2})^{2}\varphi
_{2}+i\nabla^{2}\varphi_{2}-\\
&  \left\vert \varphi_{1}\right\vert ^{2}\varphi_{1}-2\gamma\varphi_{1}%
-\gamma\varphi_{1}^{\ast}e^{i2\theta T_{2}}\nonumber
\end{align}

where $\gamma=\frac{1}{2}\frac{F^{2}}{\omega^{2}}(\omega^{2}+1)$

Finally, developing up to second order we have%
\begin{gather}
E=\varepsilon E_{1}+\varepsilon^{2}E_{2}\nonumber\\
N=\varepsilon N_{1}+\varepsilon^{2}N_{2}=\nonumber\\
\varepsilon(E_{1}-F_{1}\cos(\omega T_{1})e^{i\theta T_{2}})+\varepsilon
^{2}(E_{2}-(-i\Delta+i\nabla^{2})E_{1})=\nonumber\\
(1+(-i\Delta+i\nabla^{2}))E-F\cos(\omega t)e^{i\theta t}\nonumber\\
\varphi=\varepsilon\varphi_{1}+\varepsilon^{2}\varphi_{2}\text{
\ \ \ \ \ \ \ }\partial_{t}\varphi=\varepsilon\partial_{T_{2}}\varphi
_{1}+\varepsilon^{2}\partial_{T_{2}}\varphi_{2}%
\end{gather}

Undoing the scaling and if we make the change $\psi=i\varphi e^{i\theta T_{2}%
}$ , we can write (defining $\mu=g-1)$:%
\begin{equation}
\partial_{t}\psi=(\mu-2\gamma-i\theta)\psi-\left\vert \psi\right\vert ^{2}%
\psi+i\nabla^{2}\psi-\left(  \nabla^{2}-\Delta\right)  ^{2}\psi+\gamma
\psi^{\ast}%
\end{equation}

\bigskip


\begin{thebibliography}{99}                                                                                               %


\bibitem[1]{Cross}M.C. Cross and P.C. Hohenberg, Rev. Mod. Phys. \textbf{65},
851 (1993).

\bibitem[2]{cgr}P. Coullet, L. Gil, and F. Rocca, Opt. Commun. \textbf{73},
403 (1989).

\bibitem[3]{victorlibro}K. Staliunas and V. J. S\'{a}nchez-Morcillo,
\textit{Transverse Patterns in Nonlinear Optical Resonators} (Springer,
Berlin, 2003).

\bibitem[4]{lmn}J. Lega, J. V. Moloney, and A. C. Newell, Phys. Rev. Lett.
\textbf{73}, 2978 (1994).

\bibitem[5]{coullet2}P. Coullet and K. Emilsson, Physica D \textbf{61}, 119 (1992).

\bibitem[6]{coullet1}P. Coullet et al. Phys. Rev. Lett. \textbf{65} 1352 (1990).

\bibitem[7]{germanrock1}G. J. de Valc\'{a}rcel and K. Staliunas, Phys. Rev. E
\textbf{67,} 026604 (2003).

\bibitem[8]{germanrock2}G. J. de Valc\'{a}rcel and K. Staliunas, Phys. Rev.
Lett. \textbf{105,} 054101 (2010).

\bibitem[9]{germanrock3}G. J. de Valc\'{a}rcel, Manuel Mart\'{\i}nez-Quesada,
and K. Staliunas, Phil. Trans. R. Soc. A,\textbf{\ 372} 20140008 (2014)

\bibitem[10]{adolfo1}A. Esteban-Martin et al. Phys Rev. Lett. \textbf{94}
223903 (2005).

\bibitem[11]{kestas1}K. Staliunas, Phys. Rev. A \textbf{48 }1573 (1993).

\bibitem[12]{kestas2}K. Staliunas, M. F. H. Tarroja, G. Slekys, C.\ O. Weiss,
and L. Dambly, Phys. Rev. A \textbf{51}, 4140 (1995).

\bibitem[13]{longhi1}S. Longhi and A. Geraci, Phys. Rev. A \textbf{54}, 4581 (1996).

\bibitem[14]{longhi2}S. Longhi, Opt. Commun. \textbf{149}, 335 (1998).

\bibitem[15]{victor0}V. J. S\'{a}nchez-Morcillo, E. Rold\'{a}n, G. J. de
Valc\'{a}rcel, and K. Staliunas, Phys. Rev. A \textbf{56}, 3237 (1997).

\bibitem[16]{victor1}K. Staliunas and V\'{\i}ctor J. S\'{a}nchez Morcillo,
Phys. Lett. A \textbf{241}, 28-34 (1998).

\bibitem[17]{victor2}V\'{\i}ctor J. S\'{a}nchez Morcillo and K. Staliunas,
Phys. Rev. E \textbf{60} 6153 (1999).

\bibitem[18]{curva}R. Gallego, Maxi San Miguel and Ra\'{u}l Toral, Phys. Rev.
E \textbf{61 }2241 (2000).

\bibitem[19]{gomila}D. Gomila, P. Colet, G-L Oppo and M. San Miguel, Phys.
Rev. Lett. \textbf{87} 194101 (2001).

\bibitem[20]{taranenko}V. B. Taranenko, K. Staliunas and C. O. Weiss , Phys.
Rev. Lett. \textbf{81} 2236 (1998).
\end{thebibliography}
\end{document}